# Induced anomalous Hall effect of massive Dirac fermions in ZrTe$_5$ and HfTe$_5$ thin flakes

Yanzhao Liu,[1] Huichao Wang,[2] Huixia Fu,[3] Jun Ge,[1] Yanan Li,[1] Chuanying Xi,[4] Jinglei Zhang,[4] Jiaqiang Yan,[5] David Mandrus,[5,6] Binghai Yan,[3,*] and Jian Wang[1,7,8,†]

[1]*International Center for Quantum Materials, School of Physics, Peking University, Beijing 100871, China*

[2]*School of Physics, Sun Yat-Sen University, Guangzhou 510275, China*

[3]*Department of Condensed Matter Physics, Weizmann Institute of Science, Rehovot 7610001, Israel*

[4]*High Magnetic Field Laboratory, Chinese Academy of Sciences, Hefei 230031, Anhui, China*

[5]*Materials Science and Technology Division, Oak Ridge National Laboratory, Oak Ridge, Tennessee 37831, USA*

[6]*Department of Materials Science and Engineering, University of Tennessee, Knoxville, Tennessee 37996, USA*

[7]*CAS Center for Excellence in Topological Quantum Computation, University of Chinese Academy of Sciences, Beijing 100190, China*

[8]*Beijing Academy of Quantum Information Sciences, Beijing 100193, China*

Researches on anomalous Hall effect (AHE) have been lasting for a century to make clear the underlying physical mechanism. Generally, the AHE appears in magnetic materials, in which extrinsic process related to scattering effects and intrinsic contribution connected with Berry curvature are crucial. Recently, AHE has been counterintuitively observed in non-magnetic topological materials and attributed to the existence of Weyl points. However, the Weyl point scenario would lead to unsaturated AHE even in large magnetic fields and contradicts the saturation of AHE in several tesla (T) in experiments. In this work, we investigate the Hall effect of $ZrTe_5$ and $HfTe_5$ thin flakes in static ultrahigh magnetic fields up to 33 T. We find the AHE saturates to 55 (70) $\Omega^{-1}$ $cm^{-1}$ for $ZrTe_5$ ($HfTe_5$) thin flakes above ~ 10 T. Combining detailed magnetotransport experiments and Berry curvature calculations, we clarify that the splitting of massive Dirac bands without Weyl points can be responsible for AHE in non-magnetic topological materials $ZrTe_5$ and $HfTe_5$ thin flakes. This model can identify our thin flake samples to be weak topological insulators and serve as a new tool to probe the band structure topology in topological materials.

Anomalous Hall effect (AHE) is an important electrical transport phenomenon attracting extensive interest in both fundamental physics and potential applications [1,2]. Since the discovery of AHE in ferromagnetic iron in 1881 [3], the controversy of the microscopic mechanisms of AHE has lasted for almost a century. Two mechanisms have been identified nowadays: extrinsic process related to scattering effects and intrinsic contribution connected with Berry curvature [1,4-6]. The intrinsic AHE is quantitatively determined by the Berry curvature of the occupied states. Based on the theoretical developments, many experimental works of AHE in magnetic materials are convincingly explained by detailed band structure calculations [7-13]. Usually, time-reversal symmetry-breaking by the magnetism is taken as the prerequisite for the AHE. Counterintuitively, non-magnetic topological materials $Cd_3As_2$ and $ZrTe_5$ were recently found to show AHE or anomalous Nernst effect (ANE) [14-17] under the external magnetic field ($B$). In these materials, the Weyl points, which might be from the magnetic field induced Dirac points splitting, were believed to contribute to the AHE/ANE [18]. However, the Weyl point scenario leads to increasing AHE when the field continuously separates Weyl points, until Weyl points annihilate and generate a 3-dimensional (3D) quantized AHE, as shown in Figs. 1(a) and 1(d). This contradicts the fact that the AHE saturates quickly in the field of several tesla (T) and presents a low plateau in previous experiments (the AHE conductivity is usually tens of $\Omega^{-1}$ $cm^{-1}$) [15,17].

Transition-metal pentatelluride $ZrTe_5$ and $HfTe_5$ have been studied since the 1970s due to their outstanding thermoelectric properties [19,20]. They are predicted [21] and confirmed to be topological materials with massive Dirac bands at the border between the strong and weak topological insulators (TIs). The electronic structures of $ZrTe_5$ and $HfTe_5$ are sensitive to the interlayer coupling and lattice parameters, which makes them promising platforms to study various intriguing phenomena including log-periodic quantum oscillations [22,23], 3D quantum Hall effect [24], negative magnetoresistance (NMR) [25,26], unconventional Hall effect [27,28], etc. However, the topological categorizations of these two materials are still controversial because it is not easy for experiments to independently determine whether they are weak or strong TIs [29-31].

In $ZrTe_5$ and $HfTe_5$, clear nonlinear Hall traces are usually observed and the mechanisms for the Hall response are still under debate. The widely used two-carrier model tends to interpret the nonlinearity as the presence of more than one type of carriers [23-25,32,33]. On the other hand, Berry curvature induced AHE was claimed in the $ZrTe_5$ system, which also contributes to the nonlinearity of Hall traces [15,17]. The origin of the AHE in the non-magnetic topological materials remains to be unambiguously and quantitatively clarified. The weak interaction between the layers of $ZrTe_5$ and $HfTe_5$ allows us to obtain flakes from the bulk by exfoliation [21]. The precisely aligned Hall bar structures obtained by micro- and nanofabrication processes are advantageous for measurements compared with previous works on bulk materials. More importantly, the high magnetic field is necessary to thoroughly investigate the field dependence of AHE.

In this work, we perform systematic magnetotransport measurements on $ZrTe_5$ and $HfTe_5$ flakes with thicknesses of about 210 nm in static ultrahigh magnetic fields up to 33 T. The nonlinear Hall resistance saturates at high magnetic fields, which cannot be explained by the classical Drude model. Because the magnetic field modifies the band structure by Zeeman splitting, it can sensitively change carrier densities of two spin channels. Thus, we develop an unusual Hall model with field-dependent carrier densities and explain the nonlinear Hall trace well. The AHE saturates to 55 (70) $\Omega^{-1}$ $cm^{-1}$ for $ZrTe_5$ ($HfTe_5$) above a critical field ~ 10 T at 2 K. Our band structure calculations reveal that the nonzero Berry curvature from splitting massive Dirac bands leads to the saturated AHE, which does not necessarily require the existence of Weyl points. We note that the Berry curvature intimately originates in the Dirac bands, whereas it cannot come from ordinary bands by Zeeman splitting, as illustrated in Figs. 1(b)-1(d). Furthermore, our model reveals that the strong and weak TIs exhibit opposite signs in the field-induced AHE and identifies our $ZrTe_5$ and $HfTe_5$ thin flakes to be weak TIs.

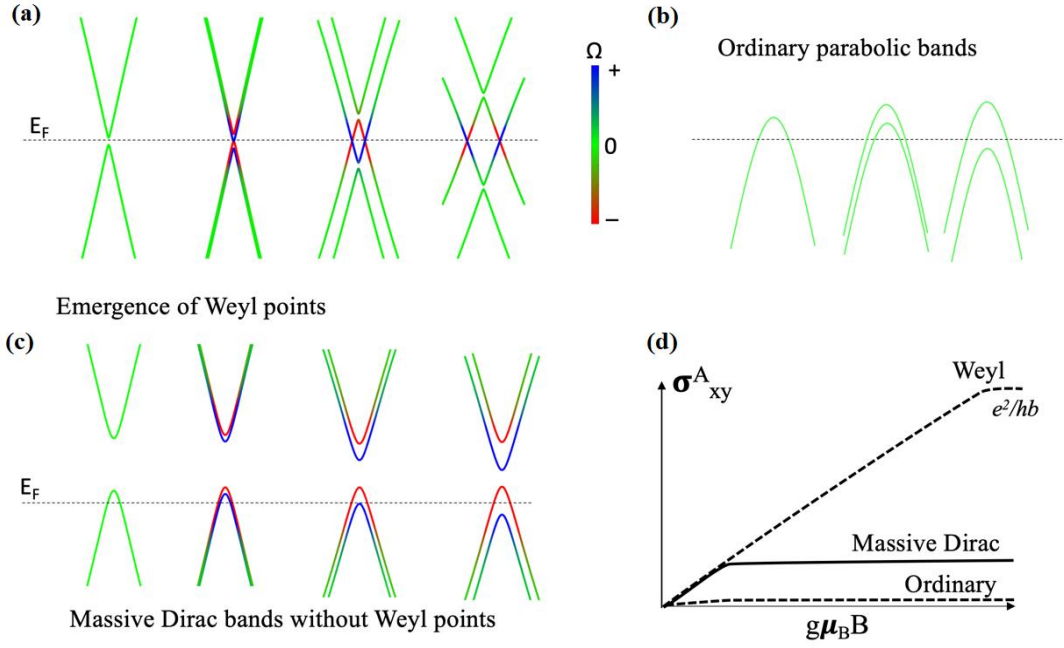

FIG. 1. Schematics of Dirac band splitting and the anomalous Hall effect. (a) Nearly massless Dirac bands split into Weyl bands as increasing the magnetic field (from the left to the right). Two Weyl points continue to split until they annihilate with each other at the boundary of two Brillouin zones, leading to 3D quantized AHE ($\sigma^A_{xy} = e^2/hb$, $b$ is the lattice parameter). Dirac bands refer to massless or small-mass Dirac states. Blue and red, respectively, represent the positive and negative Berry curvature, as shown by the color bar. (b) Ordinary bands generate negligible Berry curvature, although they also split in the magnetic field. Ordinary bands refer to common spin-degenerate bands that have negligible coupling with other bands, displaying nearly zero Berry curvature. (c) Massive Dirac bands split without generating Weyl points. An energy gap exists to gap the Dirac point and thus, two spin channels exhibit opposite, large Berry curvature near the band gap. Because the total carrier density is a constant, the Fermi surface remains unchanged after the critical Zeeman energy, leading to the saturation of the AHE. The chemical potential ($E_F$) determines the critical field. (d) Schematics of the field dependence of the AHE conductivity for three cases.

Figure 2 shows our magnetotransport results of a typical ZrTe$_5$ flake with a thickness of about 215 nm (s1). The Hall resistivity ($\rho_{yx}$) of s1 at different temperatures is represented in Fig. 2(a). The $\rho_{yx}$ grows sharply around 0 T, indicating a hole dominated carrier type ($p$ type). The slope of $\rho_{yx}$ decreases with increasing $B$, finally leading to a saturated $\rho_{yx}$. Figure 2(b) shows the magnetic field dependence of longitudinal resistivity ($\rho_{xx}$). No obvious structures appear in $\rho_{xx}$ up to 33 T, indicating that the saturation of $\rho_{yx}$ is not due to the formation of an energy gap [24]. For clarity, data curves in Figs. 2(a) and 2(b) are shifted. In addition, we show the crystal structures of the layered materials with the space group *Cmcm* in the inset of Fig. 2(a) [34]. The inset of Fig. 2(b) is the schematic of the standard six-electrode method used for electrical transport measurements. The current is applied along the *a* axis and the field is always along the *b* axis.

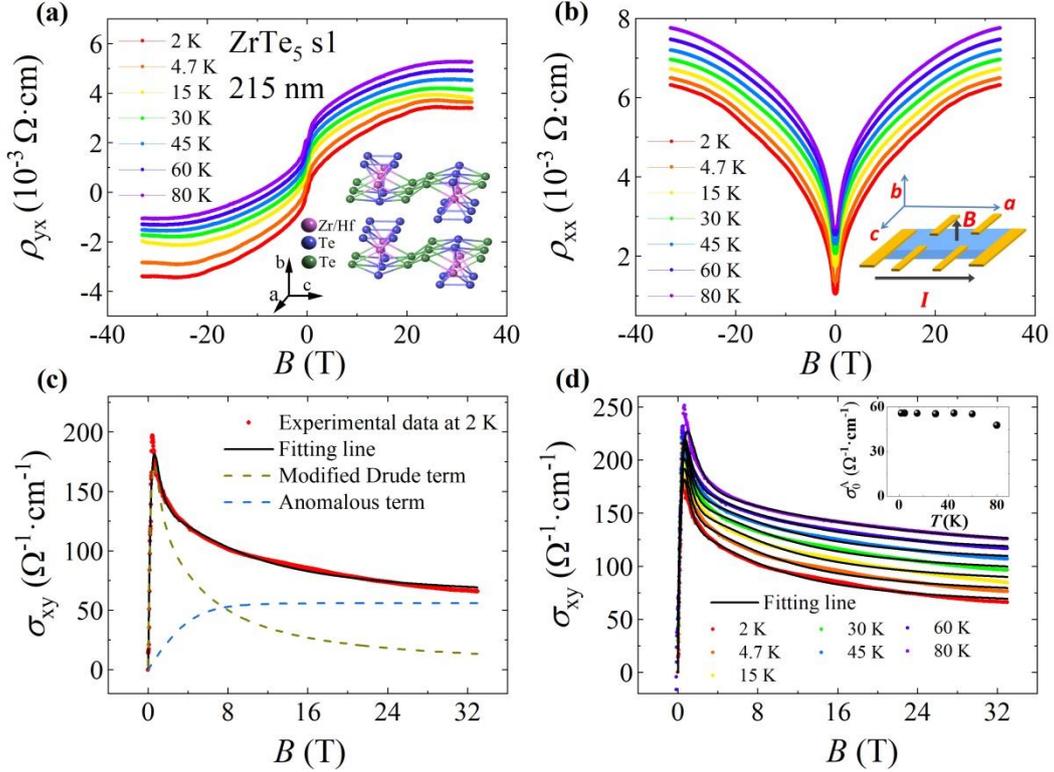

FIG. 2. Anomalous Hall effect in a ZrTe$_5$ flake (s1) with a thickness of about 215 nm. (a) Hall resistivity of s1 versus $B$ at different temperatures from 2 K to 80 K. Inset: crystal structure of ZrTe$_5$/HfTe$_5$. (b) Magnetic field dependence of longitudinal resistivity of s1 at selected temperatures from 2 K to 80 K. Inset: schematic structure for the electrical transport measurements. (c) Hall conductivity of s1 as a function of $B$ at 2 K. Red dots are experimental data; the black solid line is the fitting result by using the empirical formula based on our model, including modified Drude term and anomalous term; the dark yellow dashed line is the contribution from modified Drude model (labeled as modified Drude term) and the blue dashed line is the anomalous term. (d) Magnetic field dependence of Hall conductivity of s1 at selected temperatures. The black lines are the fitting results by using the empirical formula. The inset shows the temperature dependence of saturated values of anomalous Hall conductivity. Curves in Figs. 2(a), 2(b) and 2(d) are shifted for clarity.

We calculate the Hall conductivity ($\sigma_{xy}$) of s1 at 2 K by the relation $\sigma_{xy} = \frac{\rho_{yx}}{\rho_{yx}^2 + \rho_{xx}^2}$ and plot $\sigma_{xy}$ vs $B$ in Fig. 2(c). Because ZrTe$_5$ is a non-magnetic material and there is no contribution from the scattering of magnetic impurities to Hall response, we try to use the classical Drude model with one or two types of carriers to fit the $\sigma_{xy}$, as shown in Fig. S1(a) [35]. However, both of the fitting lines cannot match the experimental data well. The discrepancy between the Drude model and experimental data was reported in SnTe/PbTe heterostructures and Bi$_2$O$_2$Se nanoplates, which show linear magnetoresistance [39,40]. However, the physics has not been fully understood. In our observation, the unusual saturation in the Hall resistivity recalls the existence of extra anomalous Hall conductance (AHC) contribution [38]. We further adopt a different formula based on the theoretical scenario discussed in Fig. 1(c) and obtain satisfactory fitting results (black line in Fig. 2(c)) Details for the derivation of the empirical formulae are shown in the Supplemental Material [35]. The obtained empirical formulae are as follows.

$$\sigma_{xy}(B) = \sigma_{xy}^A(B) + \sigma_{xy}^N(B) \tag{1}$$

$$\sigma_{xy}^A(B) = \sigma_0^A \tanh(B/B_0) \tag{2}$$

$$\sigma_{xy}^N(B) = \left(\frac{\frac{n_0}{2}\left(1+\tanh\left(\frac{B}{B_0}\right)\right)\mu_1^2}{1+(\mu_1 B)^2} + \frac{\frac{n_0}{2}\left(1-\tanh\left(\frac{B}{B_0}\right)\right)\mu_2^2}{1+(\mu_2 B)^2}\right)eB \tag{3}$$

Here, $\sigma_{xy}^A(B)$ (blue dashed line in Fig. 2(c), labeled as the anomalous term) is the $B$-dependent AHE conductivity and $\sigma_0^A$ is the saturation value of $\sigma_{xy}^A$, $B_0$ is a parameter related to the saturation field and $\sigma_{xy}^A$ can reach $0.99\sigma_0^A$ at a critical field $B_c \sim 3B_0$, $\sigma_{xy}^N(B)$ (dark yellow dashed line in Fig. 2(c), labeled as modified Drude term) is the contribution from both carriers with spin-up or spin-down under Lorentz force, $n_0$ is the total carrier density, $\mu_{1,2}$ is the mobility of carriers with different spins. The fitting result indicates a total carrier density $n_0$ of about $2.8 \times 10^{17} cm^{-3}$ and the mobilities of the two pockets are estimated to be $1.6 \times 10^4 cm^2 \cdot V^{-1} \cdot s^{-1}$ and $3.7 \times 10^3 cm^2 \cdot V^{-1} \cdot s^{-1}$. The relatively low carrier density of the sample indicates a weak hole doping [41,42]. Figure 2(d) shows $\sigma_{xy}(B)$ at different temperatures. The temperature dependence of $\sigma_0^A$ estimated from the fitting is shown in the inset of Fig. 2(d) and the critical field $B_c$ as a function of temperature is plotted in the inset of Fig. S2 [35]. When the temperature is lower than 60 K, both $\sigma_0^A$ and $B_c$ are almost temperature-independent with values of 55 $\Omega^{-1}\cdot cm^{-1}$ and 12 T, respectively. The slight drop of $\sigma_0^A$ and enhancement of $B_c$ at high temperatures may result from the smearing effect.

To further study the universality of AHE in transition-metal pentatelluride, we carried out electrical transport measurements on a HfTe$_5$ flake with a thickness of about 205 nm (s2) at 2 K and 5 K. Figure 3(a) shows the Hall resistivity of s2 (*p* type) and the inset represents the $\rho_{xx}$ vs *B* of s2. An obvious Hall plateau can be observed at large magnetic fields, similar to those results of ZrTe$_5$ flake. The Hall conductivity of HfTe$_5$ flake cannot be fitted by the classical Drude model either (Fig. S1(b)) [35]. Meanwhile, we notice that the Eq. (3) could also well reproduce the AHC results as represented in Fig. 3(b). Curves in Figs. 3(a) and 3(b) are shifted for clarity. Inset of Fig. 3(b) shows the anomalous Hall contribution $\sigma_{xy}^A$ of s2 as a function of the magnetic field at 2 K. The total carrier density $n_0$ of s2 is $1.8\times10^{17}$ cm$^{-3}$ at 2 K. Besides, the saturation field and value of $\sigma_0^A$ are estimated to be about 8.1 T and 70 $\Omega^{-1}\cdot cm^{-1}$, respectively. The anomalous contribution in transport can also be supported by thermoelectric measurements. Figure S3 [35] represents the raw Nernst signals in a HfTe$_5$ flake with a thickness of about 210 nm (s3). A clear step-like feature with a plateau can be observed in Nernst voltages at 5 K and 10 K, consistent with the feature of ANE [14].

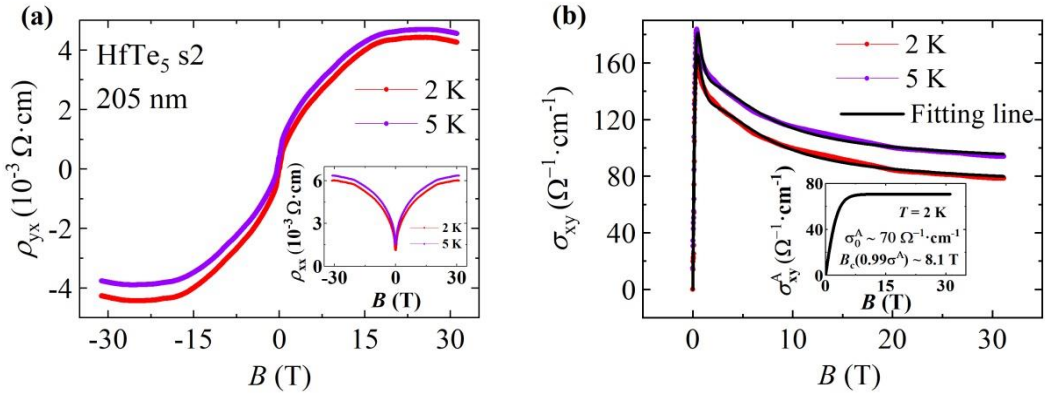

FIG. 3. Anomalous Hall effect in a HfTe$_5$ flake (s2) with a thickness of about 205 nm. (a) Hall resistivity of s2 versus *B* at 2 K and 5 K. Inset: magnetic field dependence of longitudinal resistivity at 2 K and 5 K. (b) Magnetic field dependence of Hall conductivity of s2. The black lines are fitting results based on our model. The inset shows the extracted anomalous Hall term of s2 at 2 K. The saturation field (anomalous Hall conductivity) is estimated to be about 8.1 T (70 $\Omega^{-1}\cdot cm^{-1}$). Curves in Figs. 3(a) and 3(b) are shifted for clarity.

The weak temperature dependence of the $\sigma_0^A$, as shown above, indicates the intrinsic origin of the AHE in Dirac materials ZrTe$_5$ and HfTe$_5$. Most previous works attribute the AHE to the existence of Weyl points when the conduction and valence bands cross each other by the Zeeman splitting [15]. The Weyl point mechanism gives unsaturated AHE even in large magnetic fields before the system reaches the 3D quantized AHE, which corresponds to $\sigma_0^A = \frac{e^2}{h}\frac{1}{b} \approx 260$ Ω$^{-1}$ cm$^{-1}$ ($b$ is the lattice parameter). However, the AHE saturates quickly when the magnetic field is larger than ~ 10 T (the critical field $B_c$) in our work or ~ 2 T in the previous work [15]. The $B_c$ difference may be due to the variation of the Fermi energy for different samples. Besides, the observed $\sigma_0^A$ values are one order of magnitude smaller than $\frac{e^2}{h}\frac{1}{b}$. Therefore, we can rule out the Weyl point scenario in our experiment.

Different from ordinary bands, both massive and massless Dirac bands naturally exhibit Berry curvature if the double degeneracy is lifted. Heuristically, the Berry curvature of a specific band comes from the Dirac nature, and the magnetic field can break the cancellation of Berry curvature from degenerate bands. In ZrTe$_5$ and HfTe$_5$, the large Landé $g$-factor [32,43,44] can generate sizable Zeeman splitting. As pointed out in previous calculations [21], the strong and weak TI phases depend sensitively on the lattice parameters. We build a Wannier function-based tight-binding Hamiltonian via *ab initio* density-functional theory (DFT) calculations on ZrTe$_5$. Using the DFT-relaxed lattice parameters (a=4.03 Å; b=15.00 Å; c=13.79 Å), we obtain a weak TI phase with massive Dirac bands near the Γ point and a small energy gap $E_g$ = 43 meV. Using the experimental lattice parameters (a=3.98 Å; b=14.51 Å; c=13.70 Å) [33], we obtain a strong TI phase with massive Dirac bands near the Γ point with an indirect energy gap $E_g$ = 41 meV. Then we introduce a Zeeman energy $g\mu_B B/2$ to the system and investigate the band splitting $\Delta E \approx g\mu_B B$ for both weak and strong TI scenarios.

The energy bands are doubly degenerated without the Zeeman field. The Berry curvatures of two degenerate bands cancel each other exactly. Once an external magnetic field is applied along the $b$ axis, the energy bands split into two with opposite Berry curvature. The Berry curvature distributes mainly in the band edge region, as a feature of the massive Dirac fermions. The nonzero Berry curvature on the Fermi surfaces induces the AHE since the Berry curvature from the Fermi sea is zero [6]. As bands split, the Fermi surface topology changes while the total carrier density remains the same. Therefore, we determine the new Fermi energy by fixing the total carrier density ($n_0$ in Eq. (3)) for each Zeeman energy. Then we integrate the Berry curvature over the corresponding Fermi surfaces [45] and obtain the AHE conductivity $\sigma_{xy}^A$. Based on our experiment, we set the carrier to be *p*-type and focus on the top region of the valence bands, which are located near the Γ point. We check several carrier densities from 0.4 to 10.2 $\times 10^{17}$ cm$^{-3}$ in calculations.

Figure 4(a) shows the evolution of the band structure of the weak TI phase. For increasing $g\mu_B B$, one Fermi surface expands in volume and the other shrinks until the smaller Fermi surface vanishes into a point and disappears at the critical field $B_c$. The critical field $g\mu_B B_c$ is proportional to the Fermi energy $E_F$, as shown in Fig. 4(b), where $E_F \approx g\mu_B B_c/2$ in the presence of strong spin-orbit coupling. After the critical field, the large Fermi surface remains unchanged because of the fixed total carrier density. For a very large field $g\mu_B B \approx 50$ meV, the valence and conduction bands touch each other and induce Weyl points. Because their energy is still far from the Fermi surface, the Weyl points weakly affect the Berry curvature on the Fermi surface. As shown in Fig. 4(b), the plateau feature remains nearly flat from 50 to 60 meV.

Based on our Fermi surface calculations, we find empirically $n_0 = 0.04 |E_F|^{1.77}$, where $n_0$ is in the unit of $10^{17}$ cm$^{-3}$ and $E_F$ is in the unit of meV. The order of 1.77 is between a linear dispersion (3) and a parabolic dispersion (3/2), consistent with the massive Dirac band feature. Therefore, we extract the *g*-factor independently. From Fig. 2(c), we find the critical field $B_c = 12$ T and the total carrier density $n_0 = 2.8 \times 10^{17}$ cm$^{-3}$ and obtain $E_F = -11$ meV and $g \approx \frac{2|E_F|}{\mu_B B_c} = 32$ for ZrTe$_5$ flake s1. In Fig. 3(b), we have $B_c = 8.1$ T and $n_0 = 1.8 \times 10^{17}$ cm$^{-3}$ and obtain $E_F = -8.6$ meV and $g \approx \frac{2|E_F|}{\mu_B B_c} = 37$ for HfTe$_5$ flake s2. The obtained *g*-factor is in agreement with previous measurement results [32,43,44].

We also calculate the Fermi surfaces and the AHE for the strong TI phase, as shown in Figs. 4(c) and (d). For a weak TI, both the Γ and Z points have band inversions while the strong TI phase has a band inversion only at the Z point. For the weakly *p*-doped case, the Fermi surfaces are located around the Γ point. The mass of Dirac fermions at Γ is positive (negative) for the strong (weak) TI. Therefore, strong and weak TIs exhibit opposite signs of the Berry curvature for the same Zeeman splitting, which provides a qualitative criterion to distinguish the strong and weak phases. For the *p*-type carriers, the weak and strong TIs exhibit positive and negative, respectively, signs in the AHE conductivity. If the sample is *n*-doped, the sign reverses compared to the *p*-doped case. Besides, a salient feature in the strong TI phase is the existence of Weyl points when two valence bands cross each other along the Γ-Z axis. The AHE is sensitively affected by Weyl points. Even after they are pushed below the Fermi energy after a critical field, the Weyl-point-induced Berry curvature remains on the Fermi surface. Instead of a plateau, the AHE decreases quickly after this point. Because of both the positive value of $\sigma_{xy}^A$ and the appearance of the plateau, we exclude the strong TI phase in our thin flake samples. It is noted that the theoretical $\sigma_{xy}^A$ is in the same order of magnitude as the experiments. The smallness of the theoretical value may be due to the large Dirac mass ($E_g \sim 40$ meV), which sensitively depends on the sample condition [46].

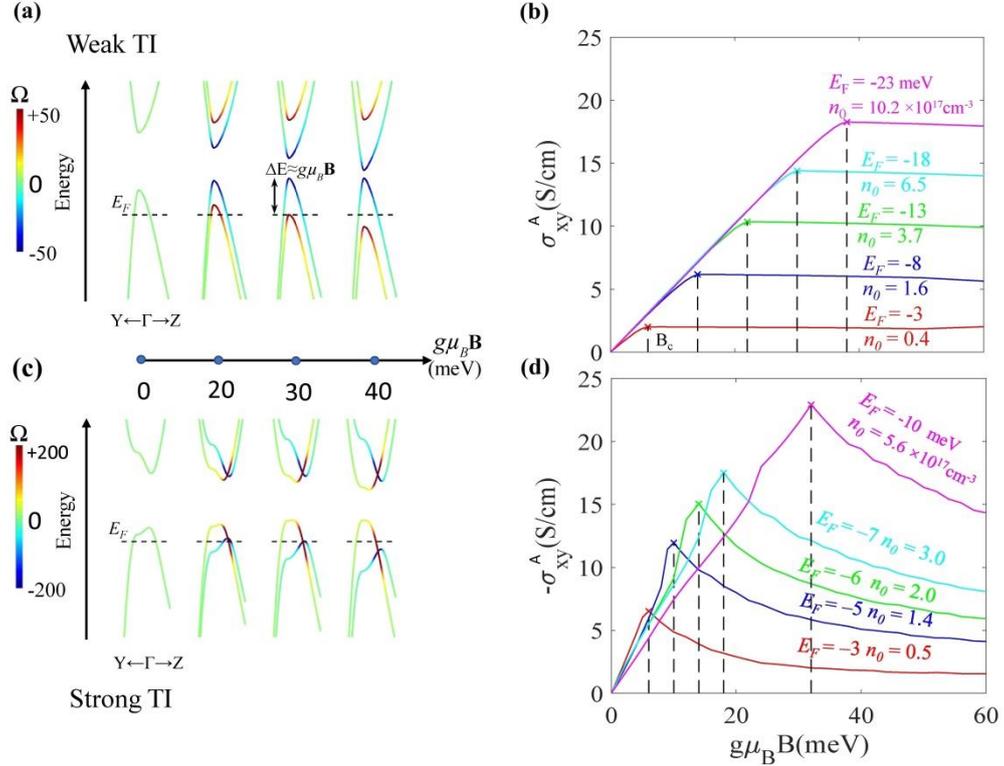

FIG. 4. The band structure and AHE evolution with respect to the Zeeman splitting. (a) The band dispersion near the Γ point for the weak TI phase. $E_F$ is the Fermi level; Δ$E$ denotes the energy splitting. (b) Anomalous Hall

conductivity ($\sigma_{xy}^A$) with different carrier densities. $\sigma_{xy}^A$ saturates after at the critical field $g\mu_B B_c$. (c)-(d) The band structure and $\sigma_{xy}^A$ for the strong TI phase. AHE exhibits an opposite sign and no plateau, compared to the weak TI.

In summary, the AHE in non-magnetic materials ZrTe$_5$ and HfTe$_5$ was clearly demonstrated via transport measurements of thin flakes under perpendicular magnetic fields up to 33 T. We reveal that the massive Dirac nature leads to the AHE in the presence of the Zeeman splitting, without involving Weyl points. The model can further distinguish the strong and weak TI phases by identifying the sign of the AHE, and our thin flake samples are characterized to be the weak TI. The results clearly clarify the AHE in transition-metal pentatellurides by high magnetic field measurements on flake samples with precise Hall bar and quantitative theoretical models. This work resolves important questions about the unusual AHE in non-magnetic topological materials and provides a useful tool to distinguish strong/weak TIs by AHE analysis.

We thank Xiaosong Wu and Wenjie Zhang for the help in related experiments. B.Y. acknowledges helpful discussions with Prof. Jiun-Haw Chu. This work was financially supported by the National Key Research and Development Program of China (2018YFA0305604 and 2017YFA0303302), the National Natural Science Foundation of China (Grant Nos. 11888101, 11774008, 12004441), Beijing Natural Science Foundation (Z180010) and the Strategic Priority Research Program of Chinese Academy of Sciences (Grant No. XDB28000000). J.Y. and D.M. were supported by the U.S. Department of Energy, Office of Science, Basic Energy Sciences, Materials Sciences and Engineering Division. H.W. acknowledges the support of the Hundreds of Talents program of Sun Yat-Sen University and the Fundamental Research Funds for the Central Universities (No. 20lgpy165). B.Y. acknowledges the financial support by the Willner Family Leadership Institute for the Weizmann Institute of Science, the Benoziyo Endowment Fund for the Advancement of Science, Ruth and Herman Albert Scholars Program for New Scientists, the European Research Council (ERC) under the European Union's Horizon 2020 research and innovation programme (ERC Consolidator Grant No. 815869, "NonlinearTopo"). C. X. was supported by the Users with Excellence Project of Hefei Science Center CAS (Grant No. 2018HSC-UE015).

Y. Liu, H. W. and H. F. contributed equally to this work.

[†] Corresponding author.

jianwangphysics@pku.edu.cn (J.W.)

[*] Corresponding author.

binghai.yan@weizmann.ac.il (B.Y.)

# Induced anomalous Hall effect of massive Dirac fermions in ZrTe$_5$ and HfTe$_5$ thin flakes

Yanzhao Liu,[1] Huichao Wang,[2] Huixia Fu,[3] Jun Ge,[1] Yanan Li,[1] Chuanying Xi,[4] Jinglei Zhang,[4] Jiaqiang Yan,[5] David Mandrus,[5,6] Binghai Yan,[3,*] and Jian Wang[1,7,8,†]

[1]*International Center for Quantum Materials, School of Physics, Peking University, Beijing 100871, China*
[2]*School of Physics, Sun Yat-Sen University, Guangzhou 510275, China*
[3]*Department of Condensed Matter Physics, Weizmann Institute of Science, Rehovot 7610001, Israel*
[4]*High Magnetic Field Laboratory, Chinese Academy of Sciences, Hefei 230031, Anhui, China*
[5]*Materials Science and Technology Division, Oak Ridge National Laboratory, Oak Ridge, Tennessee 37831, USA*
[6]*Department of Materials Science and Engineering, University of Tennessee, Knoxville, Tennessee 37996, USA*
[7]*CAS Center for Excellence in Topological Quantum Computation, University of Chinese Academy of Sciences, Beijing 100190, China*
[8]*Beijing Academy of Quantum Information Sciences, Beijing 100193, China*

[†]Corresponding author.
jianwangphysics@pku.edu.cn (J.W.)
[*]Corresponding author.
binghai.yan@weizmann.ac.il (B.Y.)

**Materials and Methods**

**Crystal details.**

The high-quality single crystals of ZrTe$_5$ and HfTe$_5$ used in this work are of the same batch as reported in our previous works [1-3]. The single crystals were grown by the Te-flux method, and characterized by powder X-ray diffraction and transmission electron microscopy. The chemical compositions of samples were checked by the energy dispersive X-ray spectroscopy, as shown in Fig. S4.

**Device fabrication.**

ZrTe$_5$/HfTe$_5$ flakes were exfoliated with scotch tape and transferred onto 300 nm-thick SiO$_2$/Si substrates. After spin-coating using poly(methyl methacrylate) (PMMA) resist, standard electron beam lithography in a FEI Helios NanoLab 600i DualBeam System was carried out. Then, an IPA: MIBK (3:1 by weight) mixture was used to develop and define the electrodes patterns. Metal electrodes (Pd/Au, 6.5/300 nm) were deposited in a LJUHV E-400L E-Beam Evaporator after Ar plasma cleaning. Finally, a lift-off process was used to remove PMMA layers with acetone. For devices used in thermal transport measurements, Pd (6.5 nm)/Au (40 nm) was deposited using e-beam evaporation to serve as a direct on-chip heater.

**Transport measurements.**

Electronic transport measurements in this work were conducted in the static magnetic field facility (33 T) in the High Magnetic Field Laboratory in Hefei. Standard six-electrode-method was used for the measurements. As shown in the inset of Fig. 2(b) and Fig. S3(a), the current and temperature gradients are applied along the *a* axis for electrical and thermoelectric measurements, respectively. The magnetic field is along the *b* axis for all measurements.

**First-principles calculations.**

We first calculated the electronic structure of bulk ZrTe$_5$ without magnetic field by using the density functional theory (DFT) code Vienna Ab initio Simulation Package (VASP) [4]. In the first-principles calculation, the projector-augmented wave pseudopotential in combination with generalized gradient approximation (GGA) exchange-correlation functional was employed to obtain the band structures. Plane waves with a kinetic energy cutoff of 230 eV were adopted as the basis set. The lattice constants of 4.03 Å×13.793 Å ×15.00 Å were chosen for the simulations. A 12 × 4 × 4 Monkhorst-Pack *k*-mesh was used for Brillouin Zone sampling. Spin-orbit coupling (SOC) was considered in all calculations. We constructed the maximally localized Wannier functions [5] from the bulk ZrTe$_5$ calculations. Then we induced Zeeman term $g\mu_B B$ in the Wannier-function-based tight-binding (WFTB) model to investigate the anomalous Hall effect under applied magnetic field.

**Detailed description of the empirical formulae**

The generation of empirical formulae shown in our main text is described as follows.

$$\sigma_{xy}(B) = \sigma_{xy}^A(B) + \sigma_{xy}^N(B) \tag{1}$$

$$\sigma_{xy}^A(B) = \sigma_0^A \tanh(B/B_0) \tag{2}$$

$$\sigma_{xy}^N(B) = \left(\frac{\frac{n_0}{2}\left(1+\tanh\left(\frac{B}{B_0}\right)\right)\mu_1^2}{1+(\mu_1 B)^2} + \frac{\frac{n_0}{2}\left(1-\tanh\left(\frac{B}{B_0}\right)\right)\mu_2^2}{1+(\mu_2 B)^2}\right)eB \tag{3}$$

First, the unusual saturation in the Hall resistivities of our ZrTe$_5$ and HfTe$_5$ flake samples recalls the existence of extra anomalous Hall conductance (AHC) contribution [6], as shown in Eq. (1).

Then the key point is how to identify the function of $\sigma_{xy}^A(B)$. Based on the theoretical estimation, we deduce that the $\sigma_{xy}^A$ increases with increasing magnetic field when the field is lower than the critical field $B_c$ and becomes saturated when the field is higher than $B_c$. This evolution is similar to the hyperbolic tangent function, which has been used to describe the anomalous Nernst effect in topological semimetals [7]. As an empirical approach, we use Eq. (2) to describe the anomalous Hall conductance.

Finally, the massive Dirac bands split under the magnetic field due to the Zeeman effect. Such splitting leads to two pockets with two spin channels. These two pockets contribute to the $\sigma_{xy}^N(B)$ under Lorentz force, shown as two Drude terms in Eq. (3). At zero magnetic field, the carrier densities of two pockets are equal. With increasing magnetic field, the density of carriers with spin-up (spin-down) increases (decreases). Eventually, carriers with spin-up become the only type of carriers that contributes to transport.

At present, it is hard to give a precise expression for the evolution of the carrier density. To reduce the complexity of the formula, we choose the similar function of $\sigma_{xy}^A$ ($\sigma_0^A \tanh(B/B_0)$) to describe the carrier density in two pockets with increasing magnetic field ($n_{1,2} = \frac{n_0}{2}\left(1 \pm \tanh\left(\frac{B}{B_0}\right)\right)$), as shown in Eq. (3). The mobilities of splitting pockets should also change with the magnetic field. However, the behavior of mobility under the magnetic field is more complicated, so we fix the mobility value as a constant in our empirical formula.

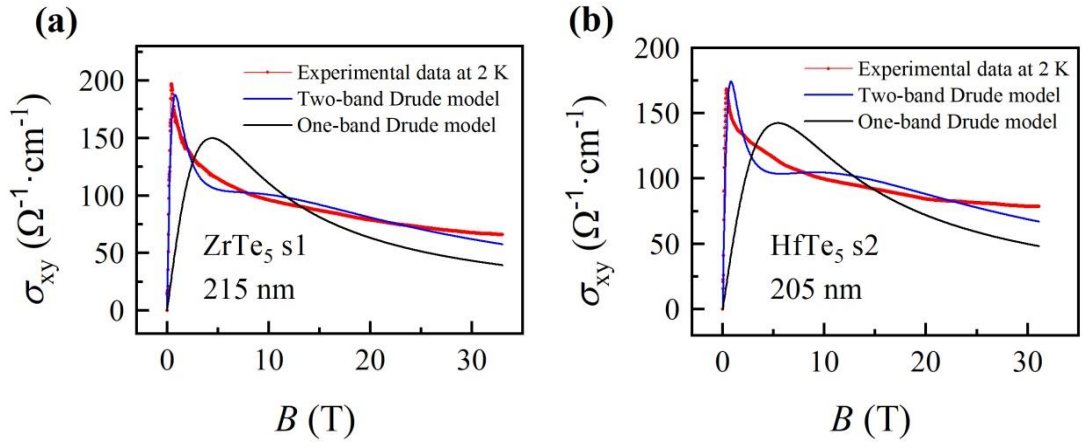

FIG. S1. Fitting results of Hall conductivity of ZrTe$_5$ flake s1 (a) and HfTe$_5$ flake s2 (b) at 2 K by using one-band and two-band Drude models.

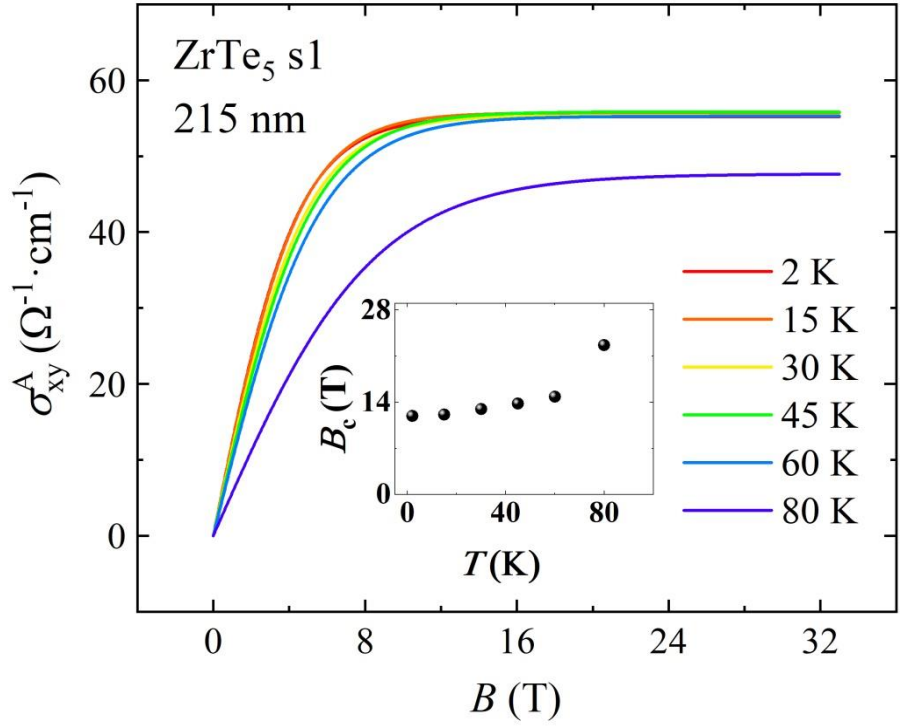

FIG. S2. Extracted anomalous Hall conductivity of ZrTe$_5$ flake s1 at selected temperatures. The inset shows the temperature dependence of the critical field.

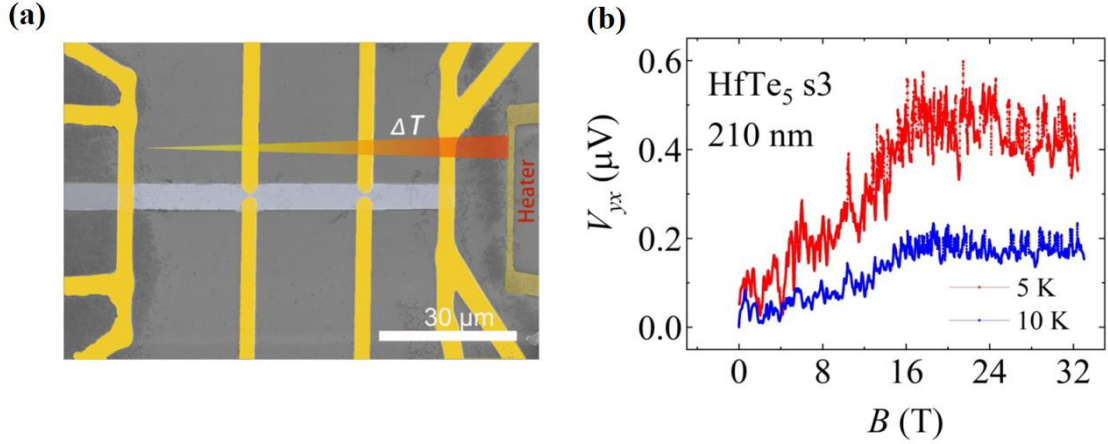

FIG. S3. Anomalous Nernst effect in a HfTe$_5$ flake (s3) with a thickness of about 210 nm. (a) Scanning electron microscope (SEM) false colour image of a typical HfTe$_5$/ZrTe$_5$ device (grey color) with electrodes (gold color). The metal stripe (pale gold color) placed on the right side of the device serves as a micro-heater. When a current passes the heater, a temperature gradient can be conducted along the $a$-axis of the sample. Scale bar represents 30 μm. (b) Nernst voltage of s3 versus $B$ at 5 K and 10 K. It is noted that the field scale of the saturation region is consistent between the raw data of Nernst voltage and Hall resistivity. The raw Nernst voltages are composed of the ordinary effect and the anomalous Nernst effect (ANE) contributed by the Berry curvature. Thus the field scale of an extracted ANE is expected to be close to that of the AHE (inset of Fig. 3(b)), given they are related to each other by the Mott's relation.

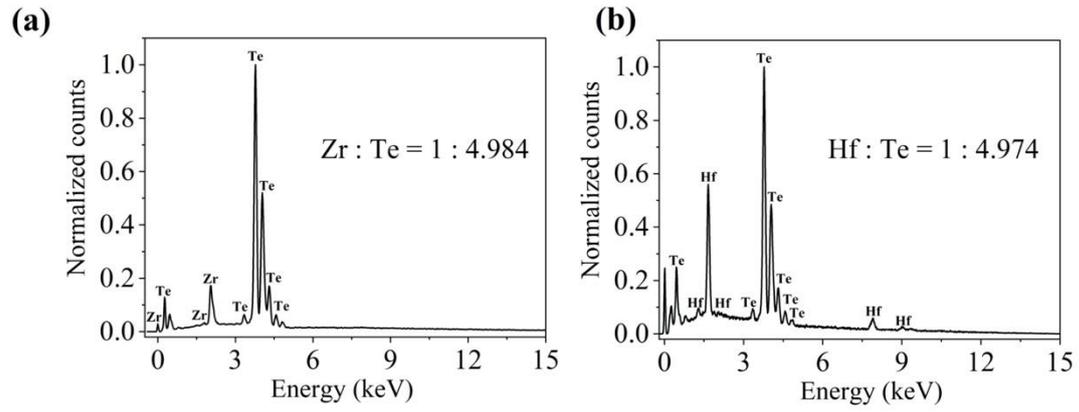

Fig. S4. Energy dispersive X-ray spectroscopy results of ZrTe$_5$ (a) and HfTe$_5$ (b) bulk crystals. The chemical composition of ZrTe$_5$(HfTe$_5$) crystals is Zr(Hf):Te ~ 1 : 4.984(4.974).

TABLE S1. Calculation parameters used in first-principles calculations.

| $E_f$ (meV) | $n$ ($10^{17}$ cm$^{-3}$) | $g\mu_B B_c$ (meV) | $\Delta E_c$ (meV) |
|---|---|---|---|
| -3.34 | 0.37 | 6 | 5.23 |
| -8.34 | 1.64 | 14 | 12.62 |
| -13.34 | 3.67 | 22 | 19.90 |
| -18.34 | 6.54 | 30 | 26.99 |
| -23.34 | 10.18 | 38 | 33.97 |